# Disinformation Echo-Chambers on Facebook

**Mathias-Felipe de-Lima-Santos [1, 2*] and Wilson Ceron [2]**

[1] Faculty of Humanities, University of Amsterdam, Science Park 904, 1098 XH Amsterdam, Netherlands
[2] Institute of Science and Technology, Federal University of São Paulo, São José dos Campos, 12231-280, Brazil
* Correspondence: m.f.delimasantos@uva.nl, mathias.felipe@unifesp.br

**Abstract**: The landscape of information has experienced significant transformations with the rapid expansion of the internet and the emergence of online social networks. Initially, there was optimism that these platforms would encourage a culture of active participation and diverse communication. However, recent events have brought to light the negative effects of social media platforms, leading to the creation of echo chambers, where users are exposed only to content that aligns with their existing beliefs. Furthermore, malicious individuals exploit these platforms to deceive people and undermine democratic processes. To gain a deeper understanding of these phenomena, this chapter introduces a computational method designed to identify coordinated inauthentic behavior within Facebook groups. The method focuses on analyzing posts, URLs, and images, revealing that certain Facebook groups engage in orchestrated campaigns. These groups simultaneously share identical content, which may expose users to repeated encounters with false or misleading narratives, effectively forming "disinformation echo chambers." This chapter concludes by discussing the theoretical and empirical implications of these findings.

## 1. Introduction

The information landscape has undergone significant transformations with the widespread adoption of the internet and online social networks. This has led to both positive and negative consequences. On the positive side, information can now spread quickly and reach a vast audience. Social media platforms have played a crucial role in fostering a culture of participation by motivating people to create and share content actively. However, there were also drawbacks. Social media platforms employ algorithms that restrict the diversity of content users are exposed to, leading to the reinforcement of pre-existing beliefs, commonly referred to as "echo chambers" [1]. These occur when individuals are exposed only to opinions that align with their own viewpoints, a phenomenon known as "confirmation bias" [2]. Furthermore, like-minded individuals often form homogeneous clusters where they reinforce and polarize their opinions through the content diffusion[3]. The functionalities and features of social media platforms, including ranking algorithms, selective exposure, and confirmation bias, have played a significant role in the development of online echo chambers [4,5].

Online social media platforms have also provided a platform for individuals with malicious intent to disseminate false or misleading narratives, with the intention of deceiving the public and undermining democratic processes [6]. The concern over false and misleading information is not a novel one. However, after the 2016 US Presidential elections[7] and the Brexit referendum in the UK [8], the propagation of extremism, hate speech, violence, and false news on platforms have significantly accentuated, highlighting their societal impact [9]. Such content often falls into the epistemological rabbit hole of "fake news" [10]. In this chapter, we use the term "disinformation" to encompass the concept of information disorder, encapsulating the so-called fake news, which includes false or misleading information created and disseminated for economic gain or intentionally deceiving the public [11].

The issue of disinformation becomes even more concerning in the context of health crises. Previous public health emergencies, such as the 2014 Ebola outbreak [12–14], showcased the widespread dissemination of inaccurate information on social media. Similarly, during the H1N1 epidemic in 2009, an array of erroneous or deceptive content propagated, ranging from conspiracy theories to unfounded rumors intended to cause harm [15,16].

Over the past decades, anti-vaccination movements have gained momentum globally [17], coinciding with a resurgence of previously controlled infectious diseases [18]. Debates on vaccines have been fueled by misleading, incorrect, and taken-out-of-context information, contributing to a perception that vaccinations are unsafe and unnecessary [18,19]. This influx of disinformation is jeopardizing the progress made against vaccine-preventable diseases, as it fuels vaccine hesitancy [20].

The COVID-19 pandemic has vividly demonstrated the disruptive potential of information disorder, shifting the focus from the health crisis toward political disinformation, which can erode the outcomes of public health policies [21]. In this context, particularly concerning are the myriad myths surrounding the safety and efficacy of COVID-19 vaccines [22–24].

Hence, online disinformation permeates all strata of society, necessitating multidisciplinary approaches for comprehension and the implementation of countermeasures. Since there is no one-size-fits-all solution to combating disinformation, experts propose a combination of interventions, such as rectifying false information and enhancing media literacy skills, to mitigate its impact [25].

Despite concerted efforts, curbing disinformation has proven to be more challenging than initially anticipated. For instance, collusive users, as outlined in the literature, purposefully promote false narratives in others' minds, amassing high counts of retweets, followers, or likes, thereby influencing public discourse. They are often funded or composed of individuals who exchange followers among themselves to amplify their visibility [26]. Such users corrode the trustworthiness and credibility of online platforms in a manner akin to spam accounts.

In response, online social media companies have adopted diverse strategies to combat information disorders. Twitter, for example, has been actively removing accounts engaged in spam and platform manipulation.

Facebook has been actively combatting problematic content on its platform since 2018 by employing the concept of "coordinated inauthentic behavior" [27]. While Facebook's efforts have been subject to criticism for their enforcement of policies, the company has substantiated the link between coordinated behavior and the dissemination of problematic information as a measure to counter manipulation attempts on its platform. This term encompasses not only bots and trolls that propagate false content but also unwitting citizens and polarized groups recruited to play orchestrated roles in influencing society.

Thus, instead of establishing a distinct demarcation between problematic and non-problematic information, the company has adopted what can be described as an "ill-defined concept of coordinated inauthentic behavior" (CIB). This strategic decision is aimed at effectively tackling and curbing the spread of disorderly information throughout its platform, all the while avoiding the complexity of unequivocally labeling it as false content [9]. Academic literature suggests that these coordinated efforts have become fertile ground for the proliferation of political disinformation [28–30], a phenomenon observed across various social media platforms [31].

This chapter undertakes an examination of disinformation narratives concerning COVID-19 vaccines that have been propagated by users on Facebook. Through the lens of echo chambers concept, this study delves into the role of user-generated content (UGC) exhibiting signs of "coordinated inauthentic behavior" within Facebook groups. To this end, the study is guided by the following research questions: (**RQ1**) To what extent, is problematic content shared on Facebook?; (**RQ2**) How are these groups interconnected?; and (**RQ3**) How do these coordinated networks possess characteristics that contribute to the formation of echo chambers?

To answer these RQs, our approach involved sourcing fact-checked stories related to COVID-19 vaccines from two major Brazilian fact-checking initiatives, namely Agência Lupa and Aos Fatos. These stories were published during the period spanning January 2020 to June 2021, and they provided the foundation for generating keywords that were then utilized in our queries on CrowdTangle. This process was instrumental in identifying false narratives that were actively circulating on Facebook. In total, our study made use of 276 instances of debunked content to uncover and analyze disinformation narratives that were being disseminated across this online social media platform. Our analysis takes the form of a computational strategy aimed at predicting instances of coordinated behavior within Facebook groups. These groups engage in inauthentic tactics with the intent of boosting the visibility and reach of particular content, ultimately contributing to the amplification of problematic information on the platform [9].

Our computational approach involves an analysis of content frequency and similarity, which enables the detection of potential traces of "coordinated inauthentic behavior." This can manifest through the replication of widely available narratives within specific Facebook groups or the sharing of common links in a condensed timeframe, often leading to external websites. Additionally, we extended our analysis to encompass the coordinated dissemination of visual content, commonly referred to as memes. These images are particularly susceptible to manipulation, rendering them more challenging to identify using conventional computational methods [32]. To address this, we leveraged a computer vision (CV) algorithm provided by Facebook to extract and analyze the textual content embedded within these images. This allowed our method to ascertain whether multiple images shared the same message over a brief period of time.

Our findings reveal a concerted endeavor to manipulate public discourse with a strategic objective of establishing "disinformation echo chambers." This is achieved by fostering a high level of engagement with false narratives across various groups, a substantial number of which are characterized by political affiliations. These fabricated information pieces possess the potential to reinforce existing biases, erode public health efforts, and trigger adverse societal consequences in relation to COVID-19 vaccines. Furthermore, the content propagated within these diverse groups can be construed as beliefs that gain potency through repeated exposure within these tightly-knit communities, effectively shielding them from counterarguments and perpetuating echo chambers [33].

In addition to these implications, the coordinated efforts to manipulate discussions within Facebook groups pose specific societal risks. This manipulation can deceive users into replicating these fabricated narratives in offline scenarios, where the tendency to resist vaccination might be exacerbated. Ultimately, our study concludes by highlighting the overarching dangers posed by these coordinated inauthentic efforts, including the propagation of confusion and mistrust among individuals, all while hindering the effectiveness of public health responses.

This chapter seeks to expand the expanding literature on disinformation and digital platforms by illustrating how coordinated inauthentic information can potentially give rise to echo chambers by effectively amplifying specific false or misleading narratives. Moreover, the scrutiny of the structural attributes of these Facebook groups, which exhibit well-defined coordinated networks, offers insights into the potential hazards and challenges posed by disinformation narratives in influencing individuals' decision-making processes regarding vaccines. This influence can result in ignorance and misperceptions that jeopardize the formulation and execution of crucial public health policies, such as vaccination campaigns [86]. The subsequent subsections delve into an exploration of the current landscape of research concerning echo chambers and disinformation.

### *1.1. Transitioning from Open Channels of Communication to Echo Chambers*

In its initial stages, online social networks were hailed for their potential to influence democracy and the public sphere by facilitating the exchange of information, ideas, and discussions in an unrestricted manner[34]. Online social media platforms embodied an optimistic perspective, driven by the disruption of traditional communication patterns in shaping public opinion, such as the gatekeeping role of newspapers to other forms of expert and non-expert communications [35]. These hopeful viewpoints championed the expansion of freedom, the transformation of democratic discourse, and the creation of a communal online knowledge hub [36]. However, these positive outlooks have given way to a more pessimistic stance, characterized by the recognition of homophily structures within these networks. This suggests that users tend to interact more frequently with individuals who share similar viewpoints, resulting in a limited range of perspectives that could foster social division and stimulate polarized outlooks [37].

Within this framework, the metaphor of echo chambers has gained prominence as a way to elucidate these behaviors amplified by the algorithms of social media platforms. It illustrates a scenario where existing beliefs are echoed and reinforced, resembling reverberations within an acoustic echo chamber [38]. Alongside the homogeneity inherent to online social networks and exacerbated by their algorithms, the concepts of selective exposure and confirmation bias have also played pivotal roles in the formation of these echo chambers within digital platforms [4,5]. Previous research has indicated that online social networks and search engines contribute to the widening ideological gap between users. Similarly, studies have identified instances of echo chambers on online social media, particularly among groups divided along ideological lines [43,44] and on controversial issues [45].

Although some studies have suggested that these effects are relatively modest [39], others argue that the term "echo chambers" might oversimplify the issue, as it is not solely a consequence of platform mechanisms but also a result of existing social and political polarizations [40]. Scholars have also put forth the argument that the extent of ideological segregation in online social media usage has been overstated, challenging the assertion that echo chambers are universally present [41].

Conversely, Facebook employs various mechanisms that could potentially exacerbate exposure to like-minded content, including the social network structure, the feed population algorithm, and users' content selection. Thus, the combination of these mechanisms might increase the exposure to ideologically diverse news and opinions. However, these mechanisms still leave individuals' choices to play a "stronger role in limiting exposure to cross-cutting content" [42].

On Twitter, researchers have examined both political and nonpolitical matters to comprehend the presence of echo chambers. According to their outcomes, political topics tend to foster more interactions among individuals with similar ideological leanings compared to nonpolitical subjects [4,38,47]. In other words, their findings suggest that homophilic clusters of users dominate online interactions on Twitter, particularly concerning political subjects [46].

In the context of studying echo chambers on online social media, it is apparent that conceptual and methodological choices significantly impact research findings [38]. For instance, studies relying on interactions or digital traces tend to indicate a higher prevalence of echo chambers and polarization compared to those focusing on content exposure or self-reported data [38]. These amplifications of pre-existing beliefs can also be shaped by the technological features of online social media platforms. In essence, the interplay between online social media interfaces and the user-technology relationship can influence the emergence of echo chambers [48].

Hence, it is crucial not only to analyze the nature of social media interactions but also to comprehend the content that users encounter in their news feeds or the groups they engage with. If the content within online groups promotes the limitation of exposure to diverse perspectives in favor of reinforcing like-minded groups that deliberately disseminate messages to larger audiences, consequently reinforcing a shared narrative, we argue that the network of groups resulting from these coordinated communication dynamics indeed resembles "echo chambers" [49]. In this chapter, we employ the term "echo chamber" to describe Facebook groups where the online media ecosystem is characterized by selective exposure, ideological segregation, and political polarization, with specific users assuming central roles in discussions.

### *1.2. The Never-Ending Challenge of "Fake News"*

Online social networks exist in a paradoxical realm, characterized by the coexistence of homophilous behavior and the potential for information dissemination. This duality has given rise to an environment where conflicting facts and contradictory expert opinions flourish, allowing false news to proliferate and conspiracies to take root [10]. Since 2016, the term "fake news" has gained global recognition as a descriptor for this false or misleading information spread in online spaces. This content can either be fabricated or intentionally manipulated to deceive individuals [11].

However, the term “fake news” has been wielded by politicians to undermine the media [10,86], leading to the emergence of alternative synonyms such as “information disorder,” “fake facts,” and “disinformation” [11]. Scholars engage in debates about differentiating between “disinformation” and “misinformation” [50]. Some argue that the distinction lies in intent, with misinformation lacking the deliberate intent to deceive. Yet, establishing intent can be challenging.

Despite these nuances, the term “disinformation” appears to be the most suitable to encompass this intricate landscape, as it covers both fabricated and intentionally manipulated content [11]. In the current complex information ecosystem, it is crucial to shift our focus from intention to the influence of the narratives that these posts align with. This is because people are not solely influenced by individual posts, but rather by the broader narratives they fit into [87]. The harmful consequences of information disorder arise from the human tendency to default to assuming the truth of a statement in the absence of compelling evidence to the contrary [5].

The rapid surge of disinformation from 2017 onward has fueled an extensive field of study, generating numerous publications approaching this multifaceted issue from diverse angles [51]. Some researchers aim to categorize various types of information disorders that emerge, while others scrutinize the social and individual dimensions of disinformation’s effects on the public and political spheres [11,52]. Computational methodologies have also been employed to detect the so-called “fake news” [51].

Over the years, automated accounts, or bots, have attracted significant attention of researchers for their potential to influence conversations, shape content distribution, and manipulate public opinion. Although terms like “bots,” “automated accounts,” “fake accounts,” and “spam accounts” have often been employed interchangeably, they do not always denote the same type of activity. *Bots* are accounts controlled by software to automate posting or interactions, while *spammers* generate unsolicited mass content. *Fake accounts*, in turn, impersonate real individuals on online platforms [53,85].

In this respect, studies demonstrate that external events and major global incidents trigger increased manipulation attempts on platforms, particularly during elections and health crises [21]. In these occurrences, traces of coordinated bot behavior could be detected in these events [9,54–56]. For example, on Twitter, estimates vary regarding the prevalence of bots, with some analyses suggesting 9% to 15% of profiles are automated accounts [57]. However, contrasting views also exist, asserting that bot accounts constitute more than 50% of Twitter users[58]. Interestingly, the platform itself provided an official statement in a public filing, indicating that fewer than 5% of its 229 million daily active users are categorized as “false” or “spam” accounts, as determined by an internal review of a sample[59].

To effectively address this issue, computational methods such as textual or social network analysis (SNA) play a crucial role in identifying and suspending harmful bots from platforms. These methods enable scholars to not only detect the detrimental effects of bots but also to mitigate their impact successfully. By understanding the nuanced differences between various types of automated accounts and their behaviors, researchers can develop more targeted strategies for preserving the authenticity and integrity of online conversations and content distribution [58].

Researchers have also identified the role of bots in amplifying the spread of disinformation and hoaxes by analyzing common interactions and network integrations. Hashtags used by these users have also been relevant for detecting automated accounts, as human users tend to use more generic ones and maintain a diverse range of social connections. Botometer, formerly known as BotOrNot, has been a widely used tool for bot detection on Twitter. It evaluates the extent to which a Twitter account exhibits characteristics similar to those of social bots, aiding in the study of inauthentic accounts and manipulation on online social media for over a decade [60]. However, scholars have also pointed out that bots are becoming more sophisticated around human behavior, which presents limitations of these tools [61]. Additionally, Botometer is exclusive to Twitter, making it challenging to detect malicious actors on other platforms.

Other techniques have been employed to detect manipulation attempts on online platforms, including disinformation and conspiracy narratives. These methodologies encompass statistical approaches like linear regression [62] as well as social network analysis (SNA) that considers the diverse relationships users form within networks. Additionally, artificial intelligence (AI) methods, such as naive Bayes models and convolutional neural networks (CNN) [63,64] have been utilized. These different techniques have been employed both individually and in combination. Despite their utility, some of these methods come with certain limitations. While AI holds potential for enhanced detection, it necessitates a wide range of input data and exhibits higher accuracy with more recent datasets. Ensuring datasets are consistently up to date is challenging. Additionally, the strategies employed by malicious bots have undergone substantial evolution in recent years, hampering these methods.

Other methods have been employed to detect manipulation attempts on platforms, including disinformation and conspiracy narratives. These techniques involve statistical methods, such as linear regression [62], social network analysis – SNA (which considers different types of relationships among users that form these networks), and artificial intelligence (AI) methods (e.g., naive Bayes models and convolutional neural networks – CNN) [63,64]. These methods have also been employed singularly or in combination. Despite this, some of them present caveats. While AI solutions hold promise for improved detection, they require diverse input data and are more accurate with recent datasets. However, datasets are not always up-to-date, and the strategies of malicious bots have evolved considerably in recent years.

Given these factors, there is a clear need for more sophisticated bot detection models or a greater reliance on methodologies that scrutinize the scope of activity within coordinated campaigns. When multiple entities collaborate within a network to achieve a common goal, the presence of coordination becomes evident [58]. In this vein, CIB strives to monitor the manipulation of information across online social networks, leveraging content dissemination through automated means to amplify its reach. This shift in focus from content and automated accounts to information dynamics within social networks aligns with Facebook's policies, which link coordinated behavior with the sharing of problematic information [9,65].

Some scholars advocate for the advancement of techniques targeting bot coordination over mere bot detection, as orchestrated bot activities can prove significantly more detrimental [58]. This aligns with Facebook's approach in its policies, employing the term CIB to underline the association between coordinated behavior and the propagation of problematic information [27].

Similarly, researchers have examined group-level features using graphs to identify orchestrated activities through users' shared relationships such as friends, hashtags, URLs, or identical messages [65]. In this respect, previous studies have explored CIB through shared links on Facebook pages, groups, and verified public profiles [9] .

Cordinated behaviors in online networks have been associated to the creation of echo chambers, as users intentionally orchestrate communication dynamics to disseminate messages to large audiences intentionally [66,67]. Another study has revealed a connection between the rapid dissemination of false information and the existence of echo chambers, primarily due to the existence of polarized clusters of opinions and networks that contribute to the spread of such information [68].

While researchers have recognized collective behavior among malicious actors driven by economic and ideological motives, the academic literature has not extensively explored coordinated mechanisms for spreading false or misleading content through messages and memes. Notably, the COVID-19 pandemic has highlighted the prevalence of visual content sharing for disseminating disinformation on online social networks [67]. In this context, Facebook groups could serve as pivotal conduits for the propagation of intricate contagions of viral disinformation.

This chapter seeks to address this knowledge gap by delving into this subject, specifically focusing on COVID-19 vaccine disinformation within public Facebook groups. In the subsequent section, we provide an in-depth overview of our methodology for pinpointing echo chambers of disinformation on the Facebook platform.

## 2. Data and Methods

### 2.1. Data Collection and Preparation

Recognizing that isolated bots might not represent the most critical issues on the platform, we opted to focus on coordinated activities to investigate disinformation campaigns within Facebook groups. We contend that these communities inherently function as echo chambers, where users intentionally join these Facebook groups to be exposed selectively to information that aligns with their beliefs and values. Hence, these communities offer an ideal context to delve into information dissemination dynamics. To explore this avenue, our study follows a three-step approach.

Initially, we identified disinformation narratives circulating on Facebook by analyzing debunked content from two prominent fact-checking agencies in Brazil: Agência Lupa and Aos Fatos. Both organizations adhere to the transparency standards set by the International Fact-Checking Network (IFCN), a coalition dedicated to upholding excellence in the fact-checking industry [69]. Our data collection spanned from January 2020 to June 2021, yielding a total of 2,860 items. We employed an algorithm to filter out debunks that did not include the term "vaccine" or related variations in their titles. This process yielded 250 debunks specifically addressing COVID-19 vaccines. Subsequently, we subjected these debunks to qualitative analysis, confirming that they were all false or misleading, and eliminating any that did not meet this criterion.

Moving on to our second step, we extracted relevant data to locate these debunked posts within Facebook. We utilized academic access to CrowdTangle, an insights tool owned and operated by Meta since 2016. It is important to note that prior research has highlighted certain limitations of this tool, such as incomplete metrics and restricted access to fully public spaces on the broader Facebook or Instagram platforms. CrowdTangle only encompasses public groups with a certain user threshold, as opposed to the entire spectrum of groups [9,70,71].

In our search, we aimed to pinpoint sentences that could be readily identified and would not yield unrelated results. For instance, we refrained from using phrases such as "COVID-19 vaccines" or similar constructs that could encompass both disinformation and credible information. Our search criteria aligned with the timeframe of the debunks, spanning from January 2020 to June 2021. Through this process, we retrieved a total of 21,614 posts containing disinformation across 3,912 groups. Importantly, this data extraction was performed after Facebook's public announcement that it had removed false content from its platform [72]. This announcement holds particular significance, as these posts should have been eradicated from the platform by that time, which could have hindered our study. Nevertheless, our findings reveal that this announcement was not fully realized, as many debunked posts persisted on the platform. This discrepancy suggests that the volume of such posts within Facebook public groups could be even more substantial.

In our third phase, we proceeded to download all the identified posts. Due to data extraction limitations within the tool, we segmented the process into timeframes, later amalgamating the data into a unified dataset. This database underwent a process of duplicate removal based on post IDs, resulting in the elimination of 1,707 duplicated entries from our initial dataset. Consequently, our final dataset encompassed 19,457 distinct entries.

### 2.2. Data Analysis and Visualization

In prior investigations of coordinated inauthentic behavior, researchers utilized estimated time thresholds to identify items shared in near-simultaneity over a short period. Similarly, a statistical metric was proposed to identify concurrent link sharing by assessing the interarrival time – the interval difference in seconds between successive shares of URLs [28]. However, we chose not to adopt these thresholds in our study for several reasons.

Unlike previous studies that centered on URLs [9,28], our analysis seeks to identify CIB within textual and visual content. Moreover, our study focuses solely on coordinated acticities among non-human accounts, necessitating a more stringent approach. This threshold determination was guided by similar studies that calculated this value based on a subset of the 10% of URLs with the shortest time intervals between the first and second shares [9,28,54–56]. Our empirical tests demonstrated that the timeframe calculated from the shortest intervals of 10% of URLs could range from 30 seconds to a minute. Consequently, depending on the dataset in use, this threshold might extend to around one minute, a timeframe that could feasibly be performed by humans.

Given these constraints, we undertook manual testing to ascertain a timeframe unlikely for consecutive human posting. Our tests indicated that this interval should be less than 30 seconds. We acknowledge that factors like internet speed and computing power might impact this performance. Nevertheless, we opted to adopt a threshold of 30 seconds between two posts, as it represented the minimum time required for consecutive posting. Our approach also considered a recursive 30-second timeframe, accounting for the possibility of repeated new posts within short intervals – a scenario unlikely to occur frequently. This approach allowed us to identify coordinated posts that were disseminated over an extended period.

Considering these temporal criteria, our computational model assessed four elements to determine coordination between posts. First, the method analyzed the "message" field, encompassing the textual content of a Facebook post. Second, it scrutinized the "description" field, which provides textual information accompanying external URLs or images shared on Facebook thumbnails. For example, the description for the post in Figure 1 was "Uma catastrófica análise sobre as vacinas contra o vírus chinês: 'Interferem diretamente no material genético'," identical to the content in the thumbnail. Third, our methodology leveraged CrowdTangle's computer vision algorithm to detect text within images and ascertain if these visual contents were disseminated through automated means. It is worth noting that prior research has highlighted that CrowdTangle's computer vision capabilities for text recognition have been a recent development and are not without limitations [71]. Lastly, our process examined whether multiple entities rapidly and consistently shared the same URL, which serves as another indicator of coordinated activity [9,28].

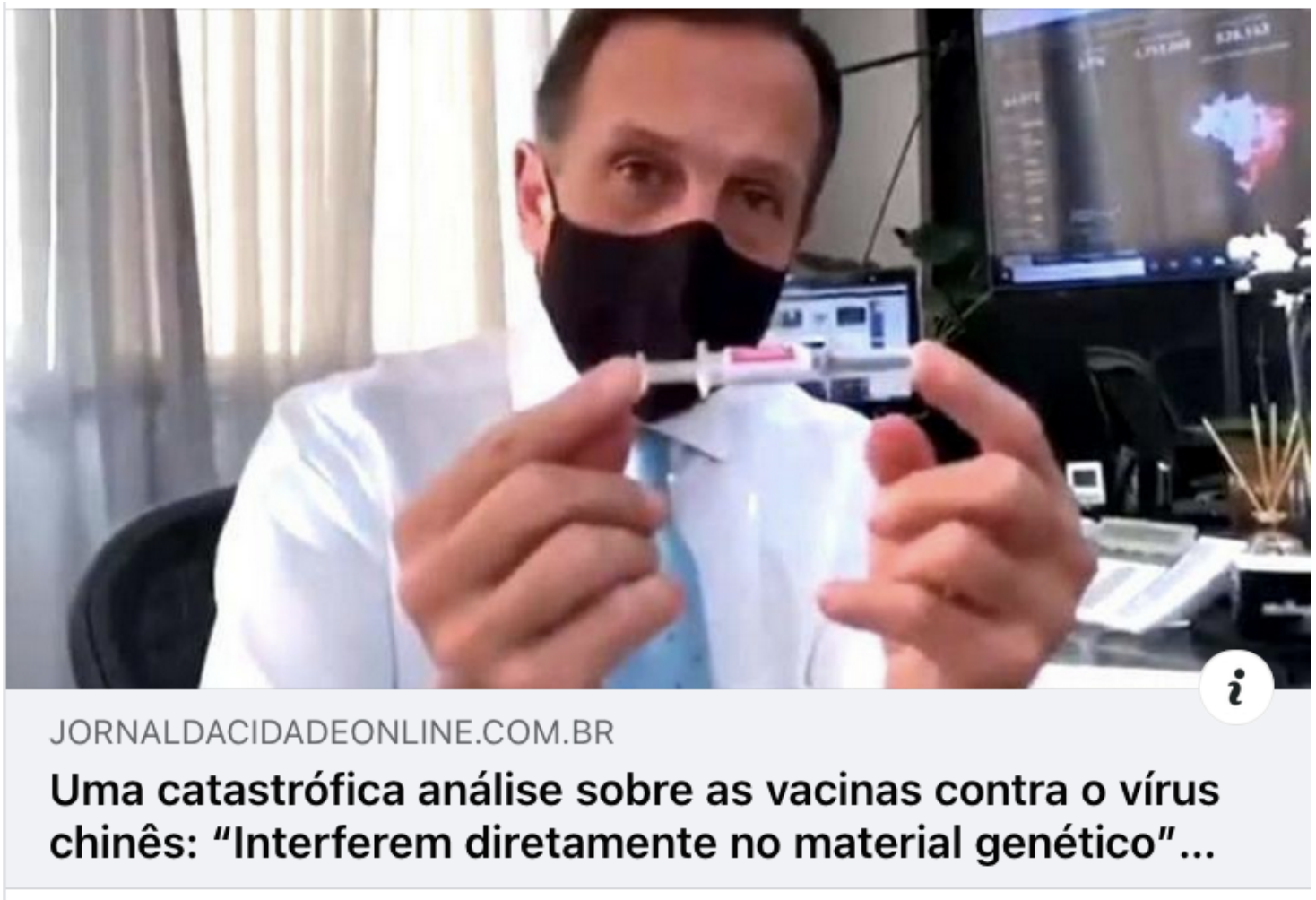

**Figure 1.** The illustration depicts a post referring to an external URL, showcased in a thumbnail. The description field is located below the link "jornaldacidadeonline.com.br." This mechanism is also observed with internal URLs, which redirect to a Facebook post.

To visualize the coordinated behaviors among different Facebook groups more effectively, we constructed a graph $G=(V,E)$, where each vertex $V=\{v_1,v_2,v_3,\ldots,v_n\}$ represents a Facebook group, and the edges $E=\{e_1,e_2,e_3,\ldots,e_m\}$ indicate the sharing of posts with signals of coordinated activity across these groups. This process was applied to the entire dataset, resulting in the creation of Figure 2. To implement this graph, we utilized the network analysis software Gephi[73], which allowed us to visually demonstrate the stronger connections between certain groups and the presence of structures that resemble "echo chambers." The Louvain method was employed to identify network communities within this graph [74]. This community detection algorithm relies on modularity optimization, resulting in a fast process to generate clusters [75]. Through this technique, we could pinpoint closely linked Facebook groups that formed more significant echo chambers.

Furthermore, we generated a second graph (see Figure 3) illustrating the five most shared instances of disinformation content. This graph, denoted as $G=(D, F)$, consisted of nodes of different types, where the set of disinformation content $D=\{d_1,d_2,d_3,\ldots,d_n\}$ was connected to the set of Facebook groups $F=\{f_1,f_2,f_3,\ldots,f_n\}$ through an edge set $E=\{e_1,e_2,e_3\ldots e_m\}$, signifying the coordinated activity signals within the dataset. This graph vividly demonstrates the robust correlation between echo chambers and the widespread dissemination of disinformation. In the subsequent section, we delve into our findings and present these visualizations.

## 3. Results

Within our dataset, we were able to identify that approximately 1,504 out of the 3,912 Facebook groups displayed indications of coordinated activity. In other words, nearly 38.5% of these groups engaged in near-simultaneous sharing of identical content. The results also underscore that these orchestrated endeavors to manipulate public discourse span across various groups with political designations. The concern is heightened considering the nature of these posts, which comprise false or misleading information.

The correlation between political Facebook groups and specific political behaviors introduces challenges to community cohesion and trust dynamics. A substantial body of literature addressing politics and social media explores the potential impact of echo chambers on individuals' behaviors and how these might undermine efforts to uphold democratic values [76,77]. These online groups, in particular, exhibit indications of selective exposure, ideological segmentation, and political polarization. In our sample, they often adopt political labels [49]. This situation compounds existing issues by occupying a privileged position in scientific communication, thereby endangering public health and hindering efforts to manage the coronavirus pandemic. These Facebook groups serve as a tangible example of the intricate and interconnected nature of disinformation rhetoric, making empirical analysis in isolation a complex endeavor. For example, past research has highlighted the penetration of political disinformation narratives in the COVID-19 discourse during the first waves of pandemic in Brazil [21].

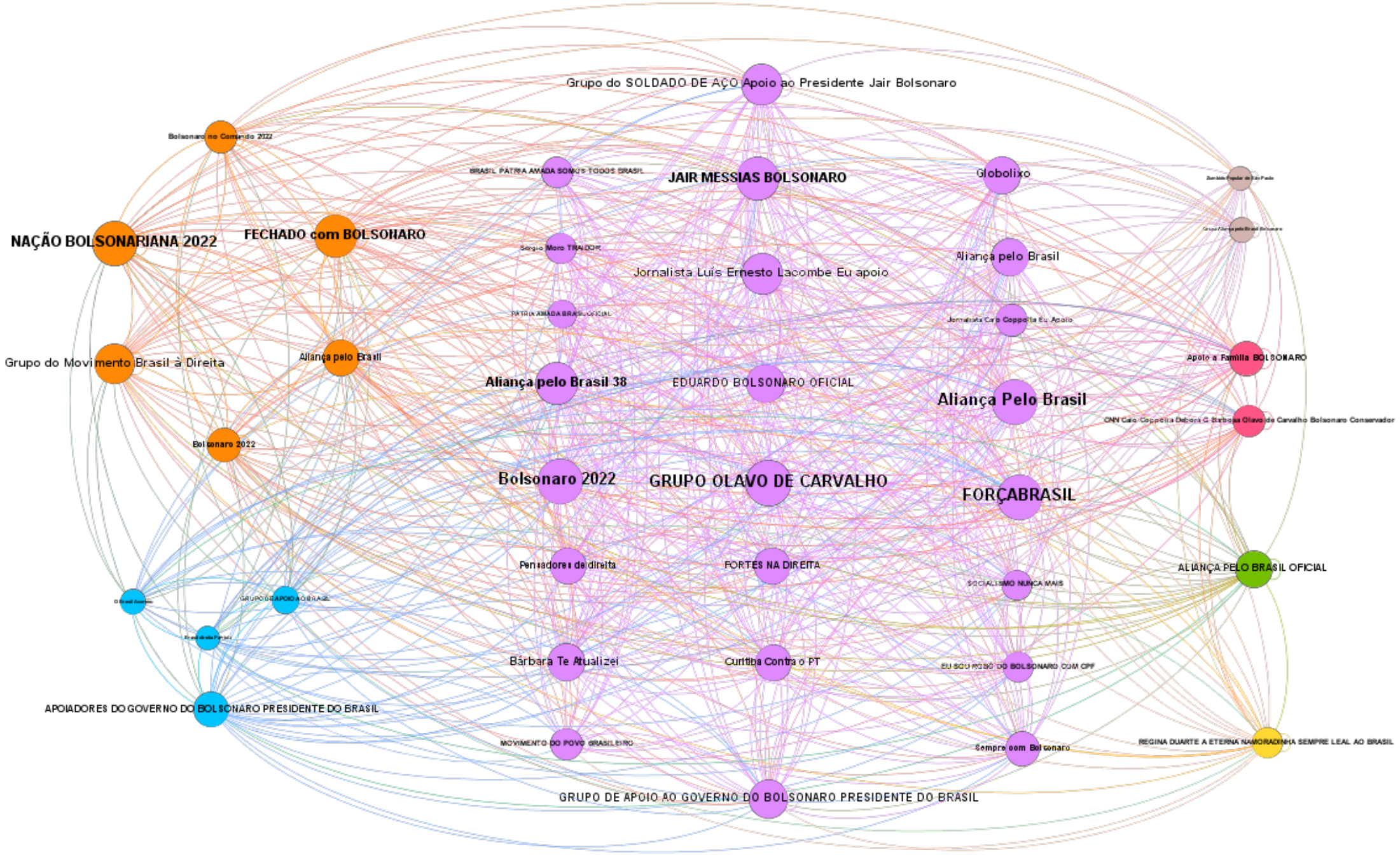

**Figure 2.** This graph exclusively features Facebook groups possessing degrees exceeding 100. In this context, these groups have shared a minimum of 100 coordinated posts. Remarkably, a significant portion of these groups have adopted political titles.

Our method successfully identified certain groups that exhibited stronger associations in disseminating these disinformation campaigns compared to others. As depicted in Figure 2, the Facebook groups highlighted in pink (a total of 117 nodes) form a particularly robust, as we called, "disinformation echo chamber." Within it, inauthentic actors appear to be swiftly and repeatedly amplifying inappropriate content. This occurrence transpires at a notably higher frequency than in other groups, as evidenced by a clustering coefficient of 0.85. This shows the propensity of nodes within this network to cluster together, resulting in the formation of triangles and the manifestation of robust community structures within this network [78,79].

Additionally, the magenta nodes consist of 257 Facebook groups that showcase coordinated behavior. These groups also exhibit a high clustering coefficient (0.83), indicating the presence of a strong community structure. Lastly, the blue nodes represent 150 Facebook groups wherein multiple actors appear to make concerted efforts to enhance the visibility of specific content by employing coordinated activities. This community boasts a more robust structure than the magenta one (with a clustering coefficient of 0.84), albeit with a smaller number of nodes. Our analysis further revealed the existence of smaller communities that also display traces of activities aimed at artificially boosting the popularity of certain online content. Consequently, these Facebook groups, which likely emerge from orchestrated communication dynamics intending to disseminate messages to wide audiences, can be likened to "disinformation echo chambers."

Figure 3 underscores how the five most frequently shared narratives are extensively propagated among these Facebook groups. Housing potentially inauthentic actors, these online communities appear to amplify these problematic contents in an endeavor to elevate their visibility. This creates a causal connection that potentially links the spread of disinformation with the presence of online echo chambers [68].

In essence, when a network of groups within an online media environment engages in nearly simultaneous and recurrent sharing of disinformation narratives, the emergence of "disinformation echo chambers" becomes apparent.

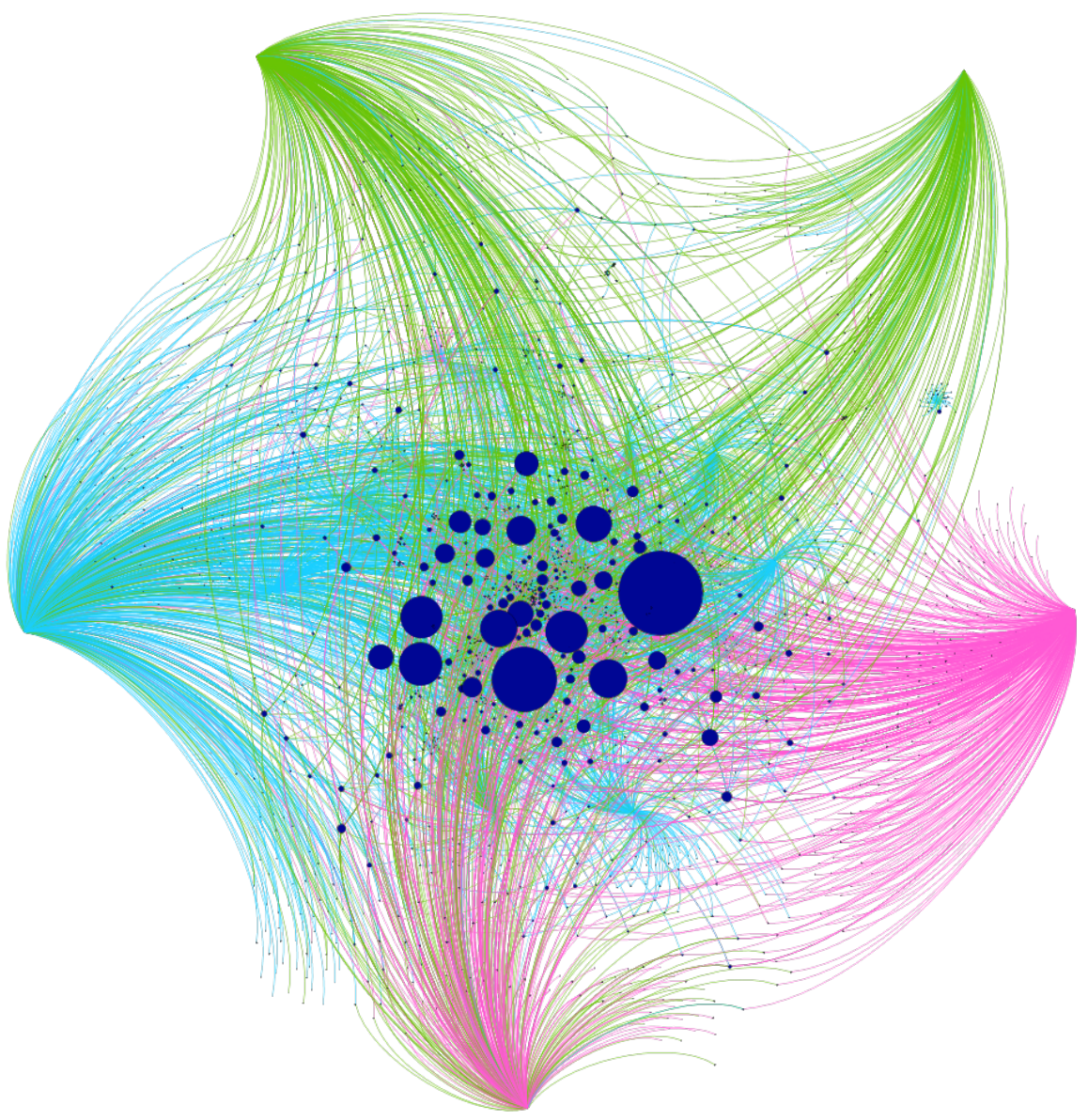

**Figure 3.** This graph illustrates the five most widely shared instances of disinformation content, highlighting their interconnectedness across various groups (dark blue nodes at the center). The edges portrayed in pink signify the shared videos within these groups, while the blue edges represent memes/photos, and the green edges signify URLs.

## 4. Discussion and Conclusions

This chapter delves into the concerning prevalence of digital disinformation within online social networks, specifically highlighting how political Facebook groups have become conduits for amplifying the reach of such narratives. Our approach successfully identified instances of disinformation narratives being shared in close proximity by various entities within a short timeframe. This encompassed URLs, posts, and memes, all of which contributed to the proliferation of echo chambers on online social media platforms.

In fact, Facebook groups inherently function as echo chambers, as users deliberately join these groups to expose themselves selectively to information that aligns with their pre-existing beliefs and values [38]. However, these groups reinforce confirmation biases and contribute to the polarization of views by limiting exposure to diverse perspectives. These Facebook groups are the spaces where one gets their daily dose of confirmation bias, exacerbating their problematic behavior [4,5]. Although some of these habits are influenced by both social and political polarization as well as platforms' algorithms, Facebook groups have emerged as fertile ground for disseminating false or misleading information [80]. This is especially evident during periods of uncertainty such as the COVID-19 pandemic [22].

Our research outcomes highlight the significant purposefully interconnection among particular groups, driven by coordinated endeavors to propagate disinformation narratives. Specifically, these posts, especially those linking COVID-19 vaccines with inaccurate or deceptive information, have played a role in fostering the expansion of anti-vaccination sentiments. This network of interrelated groups, united by the circulation of shared content, underscores the echo chamber phenomenon, wherein they reinforce their confirmation biases. Consequently, it is plausible to view these Facebook groups as "disinformation echo chambers."

We assert that these "disinformation echo chambers" emerge from orchestrated actions aimed at intentionally spreading false or deceptive narratives to wide audiences. In our context, this poses threats to strategies aimed at curbing the impact of the COVID-19 pandemic, including vaccination efforts [23,24]. Furthermore, our study underscores that, despite efforts to eliminate false or misleading content related to COVID-19 vaccines, such material remained accessible to users, even when it had been debunked by fact-checking organizations collaborating with Meta/Facebook. This situation is concerning, as it indicates that the effectiveness of these measures is questionable.

It is crucial to note that our analysis primarily focused on coordinated activities driven by automated accounts. However, real users can also contribute to coordinated inauthentic behavior [81], as recently highlighted by Facebook's expanded policies against such actions. The company announced a crackdown on coordinated campaigns of actual users that cause harm on and off its platforms, expanding its measure against coordinated activities [82].

Our study's fixed threshold approach might not capture all instances of near-simultaneous sharing, considering the evolving strategies of malicious actors. Addressing such complex scenarios requires combining various methods and approaches to effectively combat information disorder in rapidly changing online environments.

Similarly, our analysis was limited to large public groups. Similar dynamics might be at play in smaller and private groups, potentially exacerbating exposure to false narratives to these individuals. Exploring the interplay between false content dissemination in private and public groups could be a fruitful avenue for future research.

In conclusion, the ongoing pandemic has underscored the critical importance of comprehending and countering the propagation of problematic information online. This study presents an innovative computational method that uncovers the existence of "disinformation echo chambers" within public

Facebook groups using different ways of manipulate the public discourse (e..g, memes, URLs, etc.). By disseminating deceptive narratives, these groups can undermine COVID-19 vaccination efforts and erode public trust in health measures. Our findings not only shed light on these inauthentic tactics but also suggest novel approaches for detection and mitigation to combat the visibility and impact of misleading content.

**Funding**

This project was partially funded by the University of Amsterdam's RPA Human(e) AI and by the European Union's Horizon 2020 research and innovation programs No 951911 (AI4Media).